%% file: main.tex
\documentclass[aps,prd,twocolumn,superscriptaddress,nofootinbib,floatfix]{revtex4-2}
\usepackage{graphicx}
\usepackage{amsmath}
\usepackage{xurl}
\usepackage[colorlinks=true, allcolors=blue]{hyperref}
\usepackage{siunitx}
\usepackage{multirow}
\usepackage{booktabs}
\usepackage{dcolumn}
\usepackage{bm}
\usepackage{lineno}
\usepackage{float}

\usepackage{orcidlink} 
\usepackage{currfile}

\usepackage{needspace}

\begin{document}
\hyphenpenalty=10000
\exhyphenpenalty=10000
\widowpenalty=10000
\clubpenalty=10000

\title{Direct Measurement of the $^{212}\mathrm{Pb}$ and $^{214}\mathrm{Pb}$ $\beta$ Decay Branching Ratios with the XENONnT Experiment}

\input{affiliations} 
\input{authors}
\author{(The XENON Collaboration)}

\begin{abstract}
We present precision measurements of $^{212}\mathrm{Pb}$ and $^{214}\mathrm{Pb}$ $\beta$ decay branching ratios using $^{220}\mathrm{Rn}$ and $^{222}\mathrm{Rn}$ calibration data from the XENONnT detector, a dual-phase liquid xenon time projection chamber. Characterizing these isotopes is critical, as they lead to significant low-energy backgrounds in rare-event searches. We report ground-state branching ratios of $(14.75 \pm 0.20(\mathrm{stat}) ^{+0.14}_{-0.40}(\mathrm{sys}))\%$ for $^{212}\mathrm{Pb}$ and $(9.8 \pm 0.3(\mathrm{stat}) ^{+0.8}_{-0.2}(\mathrm{sys}))\%$ for $^{214}\mathrm{Pb}$, providing the most precise direct measurements of these transitions to date. These results contribute to enhancing background modeling for dark matter and neutrino experiments, improving sensitivity to solar neutrinos and physics beyond the Standard Model.
\end{abstract}

\maketitle

\section{Introduction}

Radon radioactivity is a major background source in present and future experiments utilizing liquid xenon (LXe) Time Projection Chambers (TPCs) searching for signals such as dark matter~\cite{xenon_nt_wimp_II, lz_wimp_II, panda_wimp_II} and solar neutrino scattering~\cite{DARWIN_pp_I, DARWIN_pp_II, XLZD_design_book, panda_pp}. Accurate interpretation of experimental data and the distinction between new physics and known processes require a comprehensive understanding of these isotope decays. The $^{220}\mathrm{Rn}$ and $^{222}\mathrm{Rn}$ radioactive chains include the $\beta$ decays of $^{212}\mathrm{Pb}$ and $^{214}\mathrm{Pb}$, respectively. Details of ground-state (GS) decay are essential as they directly impact the region of interest for low-energy rare-event searches. This is particularly relevant for $^{214}\mathrm{Pb}$, which currently constitutes the dominant background in XENONnT~\cite{XENONnTLowERSR0}.

\begin{figure}
    \centering
    \includegraphics[width=0.9\linewidth]{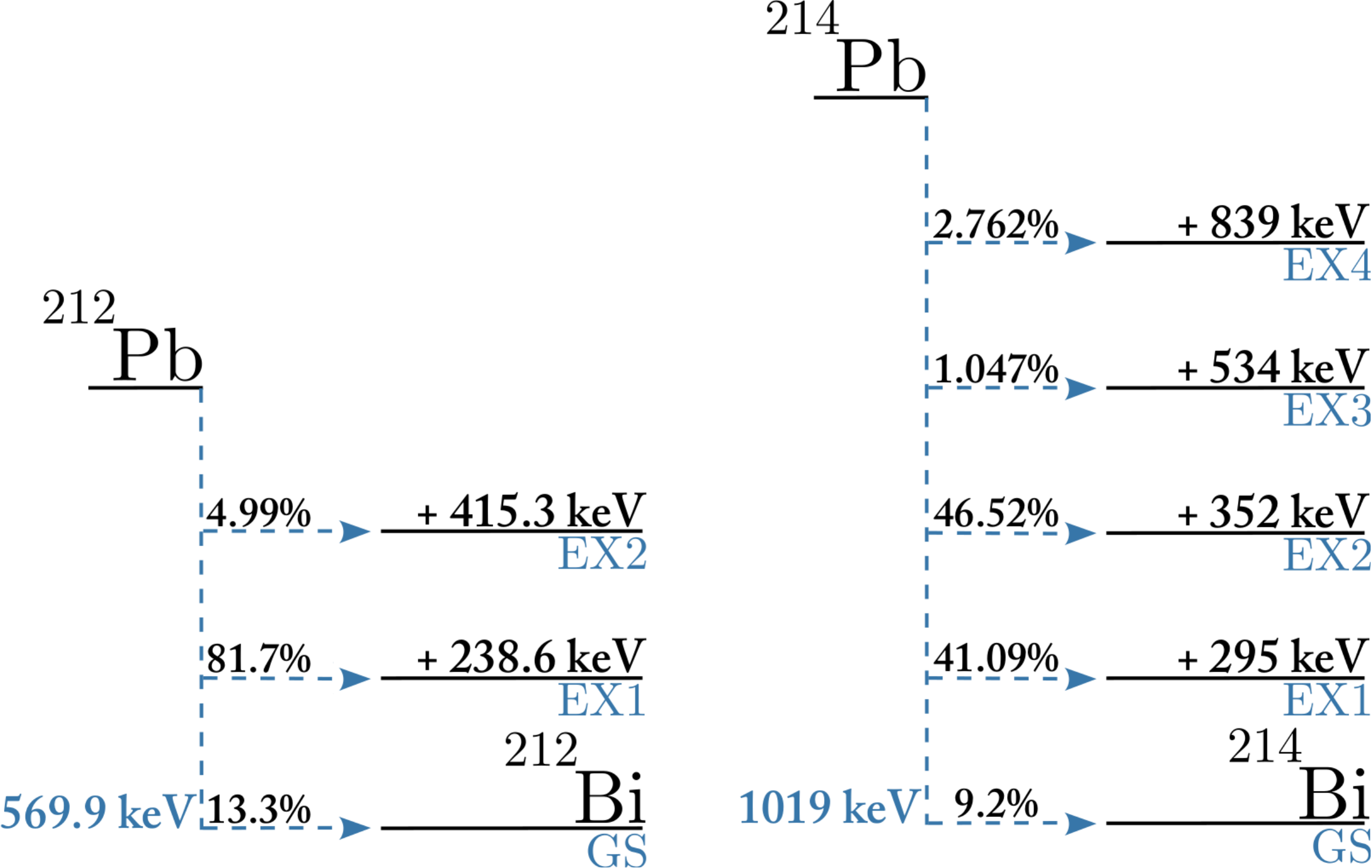}
    \caption{Decay schemes of $^{212}\mathrm{Pb}$ and $^{214}\mathrm{Pb}$. The branching ratios, Q-values and the excited state energy of the daughter nuclides are from LNHB~\cite{Be2016Table}. The $^{214}\mathrm{Pb}$ decay branch to 797\,keV $^{214}\mathrm{Bi}$ excited state ($\text{BR}\sim0.0196$\,\%), is not shown.}
    \label{fig:decay_scheme} 
\end{figure}

The main $\beta$ decay branches for both $^{212}\mathrm{Pb}$ and $^{214}\mathrm{Pb}$ are classified as first-forbidden non-unique transitions, whose spectral shape has been studied extensively but remains a matter of investigation~\cite{Mougeot, Ramalho}. As shown in Fig.~\ref{fig:decay_scheme}, the resulting $^{212}\mathrm{Bi}$ and $^{214}\mathrm{Bi}$ daughter nuclei are predominantly produced in an excited state, which then promptly de-excites by emitting one or more $\gamma$ rays. The information reported in Fig.~\ref{fig:decay_scheme} are from the Laboratoire National Henri Becquerel\,(LNHB), a French national metrology laboratory~\cite{Be2016Table}. Another common reference is the Nuclear Data Sheet (NDS)~\cite{AURANEN2020117, ZHU20211}.
The branching ratios (BRs) for $^{212}\mathrm{Pb}$ were indirectly determined from $\gamma$-ray spectroscopy and internal conversion intensities and have not been updated since the 1990s. A direct measurement of the GS branching ratio would significantly reduce its current relative uncertainty at approximately 7.7\,\%~\cite{Ferrari2025Radon}. For $^{214}\mathrm{Pb}$, the BRs are less certain. The absolute precision for the GS transition is 7–10\,\%, but recommended literature values exhibit notable tension: LNHB reports $(9.2 \pm 0.6)$\,\% while NDS reports $(12.7 \pm 1.1)$\,\%~\cite{Be2016Table, ZHU20211}. This discrepancy can be attributed to the high $\gamma$-ray multiplicity of the decay, and it has prompted multiple re-evaluations of the decay data~\cite{Ferrari2025Radon}.

The XENONnT experiment offers a novel approach to measuring BRs by modeling the complete $\beta + \gamma$ signal and the background, rather than relying only on $\gamma$-ray de-excitation. 
We report the first direct, high-precision measurements of the $^{214}\mathrm{Pb}$ $\beta$ decay BRs using XENONnT $^{222}\mathrm{Rn}$ calibration data, which addresses existing tensions in the available nuclear data. We extend this same analysis technique to $^{220}\mathrm{Rn}$ calibration data from XENONnT to determine the $\beta$ decay BRs for $^{212}\mathrm{Pb}$. 

\section{The XENONnT detector and topology of $^{212}\mathrm{Pb}$ and $^{214}\mathrm{Pb}$ $\beta$ Decays}

The core detector of the XENONnT experiment, located at the underground laboratory of Laboratori Nazionali del Gran Sasso (LNGS) in Italy, is a dual-phase TPC containing a 5.9\,t active LXe target. The cylindrical TPC, with a height of 148.6\,cm and a radius of 66.4\,cm, is equipped with 494 3-inch Hamamatsu R11410-21 photomultiplier tubes (PMTs), divided into two arrays at the top and bottom. A cathode electrode at the bottom of the TPC, a gate electrode at the top, and an anode electrode positioned a few millimeters above the gate establish the electric drift field and the extraction drift field. Background mitigation is achieved through two active veto detectors surrounding the TPC. Continuous xenon purification in both the liquid and gas phases removes electronegative impurities, and cryogenic distillation reduces radioactive contaminants. A more detailed description of the XENONnT detector and its subsystems is available in the dedicated publication~\cite{nT}.

The energy transferred from a particle interaction in the target excites and ionizes the xenon atoms. The de-excitation and electron-ion recombination generate the prompt, primary scintillation signal (S1). The ionization electrons that survive recombination are drifted upward and extracted into the gas phase, where they generate proportional scintillation (S2) via electroluminescence. The deposited energy, the interaction vertex, and the nature of the interaction can be reconstructed using the S1 and S2 signals. 
Based on the multiplicity of S2s in an event, it is possible to distinguish single-site (SS) from multiple-site (MS) events. Resolving multiplicity, of primary importance in dark matter interaction searches, has also been found to be suitable for $\beta + \gamma$ spectroscopy, as shown in this work and other XENONnT studies~\cite{b3r7-6ff4, Ferrari2025Radon, Volta2023Characterization}. 

Electrons with energies below the MeV scale in LXe have a range of ($\mathcal{O}(1\,\text{mm})$~\cite{NIST_StoppingPower}) that is smaller than the detector's spatial resolution ($\mathcal{O}(1\,\text{cm})$)~\cite{Volta2023Characterization}. Consequently, $\mathrm{Pb}$ transitions to the $\mathrm{Bi}$ nuclear GS appear exclusively as SS events. In contrast, decays to excited states emit prompt de-excitation $\gamma$ rays that can either merge into a single reconstructed energy deposit (an SS topology) or be reconstructed as separate electron and $\gamma$-ray S2 signals (an MS topology). The ability to separately reconstruct these energy depositions depends strongly on the interaction depth within the TPC, with deeper events being harder to resolve, and whether the de-excitation $\gamma$ ray was emitted parallel to the drift field.

\section{Data Selection and Efficiency}

The analysis of $^{212}\mathrm{Pb}$ decay was performed using data from the $^{220}\mathrm{Rn}$ calibration campaign conducted during the first XENONnT science run (SR0), whereas the $^{214}\mathrm{Pb}$ analysis utilized data from a novel $^{222}\mathrm{Rn}$ calibration campaign in the second science run (SR1). During these calibrations, the gaseous Xe flow is exposed to implanted $^{226}\mathrm{Ra}$ and $^{228}\mathrm{Th}$ sources~\cite{J_rg_2023, Jrg2022ProductionAC}, the progenitors of the two radon decay chains. The emanated Rn atoms are flushed into the TPC and thoroughly mixed within the volume. Their subsequent decay produces a homogeneous distribution of $\mathrm{Pb}$ isotopes, yielding high-statistics datasets that exceed the typical background rate by at least two orders of magnitude. Notably, the $^{222}\mathrm{Rn}$ calibration was possible because XENONnT is equipped with a radon removal system capable of quickly restoring $^{222}\mathrm{Rn}$ activity to pre-calibration levels~\cite{radonremovalxenonntsolar}. No impact from long-term contamination, such as increased activity of $^{210}\mathrm{Pb}$ on PTFE walls, was observed.

The events were reconstructed using the standard XENONnT data processing pipeline~\cite{strax, straxen, XENONnT_AP1}. We selected events where at least 3 PMTs contributed to the S1 signal, with an S2 threshold of $>$500\,PE. A series of data selection criteria were then applied to minimize non-physical or mis-reconstructed events and reduce other background sources. We employed a subset of the selection rules used in the low-energy electronic recoil physics search~\cite{XENONnTLowERSR0}, those for which the parameter space could be accurately simulated in the extended energy region.

Events were required to be SS, characterized by the absence of additional S2-like signal in the event waveform. We decided to exclude MS events in favor of a cleaner signal selection. Events exhibiting unusually large S2 pulse width were rejected, as well as events where a single PMT contributed the majority of the total S1 area. For the $^{212}$Pb analysis, additional selection criteria were applied based on the presence of additional S1 signals in the waveform, the score of a Naive Bayes classifier~\cite{XENONCollaboration:2023dar}, and the fraction of S2 light observed by the top array. The variation in data selection between the two analyses stems from the differing maturity levels of the SR0 and SR1 pipelines at the time of the $^{214}\mathrm{Pb}$ study. 

\begin{figure}
    \centering
    \includegraphics[width=0.9\linewidth]{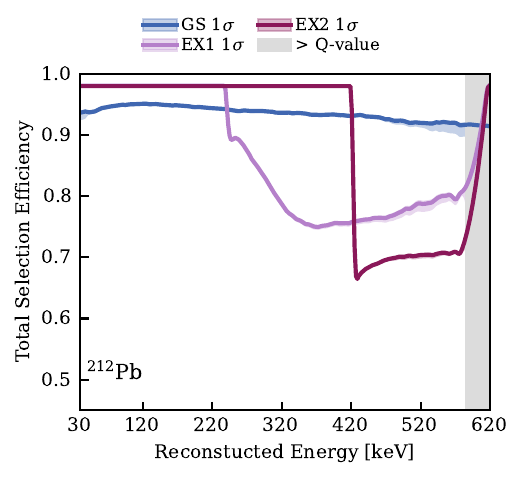}
    \includegraphics[width=0.9\linewidth]{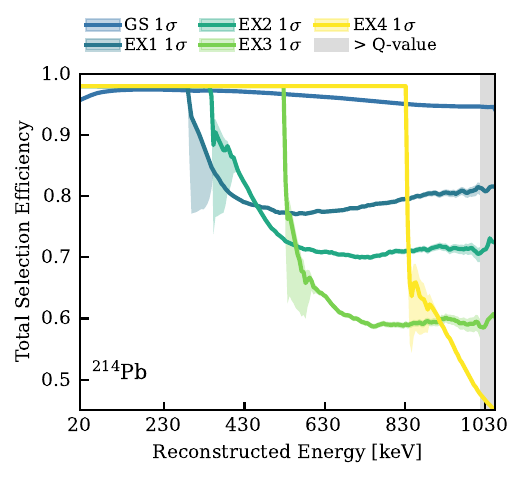}
    \caption{Total selection efficiency as a function of reconstructed energy for $^{212}\mathrm{Pb}$ (top) and $^{214}\mathrm{Pb}$ (bottom) in the used fit region. Solid lines and shaded bands represent the mean and $1\sigma$ uncertainty, respectively, for ground-state (GS) and excited-state (EX$n$) transitions. Lower efficiencies for EX$n$ reflect the impact of selections on MS topologies compared to SS GS events. Background acceptance follows the GS curves. Gray bands indicate the region above the Q-values.}
    \label{fig:eff} 
\end{figure}

To account for specific $\beta+\gamma$ event topologies, the impact of data selection was evaluated independently for each excited state. We used the Geant4 toolkit and \textit{G4RadioactiveDecayPhysics} list~\cite{ALLISON2016186, AGOSTINELLI2003250} to simulate these event topologies, modifying the radioactive decay chains to isolate individual $\mathrm{Pb} \rightarrow \mathrm{Bi}$ decay branches and ensure that efficiencies could be determined for each channel. The energy depositions were then processed through the XENONnT waveform simulator~\cite{peter_gaemers_2022_7216324, Ramirez:2022}, which models liquid xenon microphysics and detector effects to produce realistic PMT signals. This allowed the simulated data to be processed using the same XENONnT software, exploited for the detector recorded data. The only data selection efficiency estimated using a data-driven approach was the one based on the fraction of S2 light in the top array. In energy regions where no simulation data were generated, the efficiency was set to unity. Notably, the selection efficiency for excited states results suppressed relative to the GS, driven primarily by the S2 pulse width selection. To smooth the curves and avoid introducing features caused by statistical fluctuations, a running average and spline interpolation were applied. To obtain the total signal and background efficiencies, the data-driven detection efficiency and S2 signal threshold acceptance were also included. Figure~\ref{fig:eff} shows the efficiency curves for all signal components as a function of reconstructed energy.
The sharper behavior shown at the lower threshold of the EX2 of $^{212}$Pb is specifically introduced by the event selection based on S1 pulse shape.

To compute the total selection efficiency for the backgrounds featuring the $\beta+\gamma$ topology, such as the lead plated out on PTFE and the bismuth skin in the next Section described, the very same methodology was used.
For the SS-only background components, on the other hand, since they share the same topology as the GS ones, we assumed the same curve used for the GS transitions. 
To validate the method, the SS-to-MS ratio and the overall data quality acceptance (i.e., where all transitions are considered simultaneously) as a function of reconstructed energy were compared between data and simulation. The agreement resulted to be within a few percent, and this slight discrepancy is propagated as a systematic uncertainty.

A fiducial volume (FV) selection of $(3.71 \pm 0.12)$~t was applied based on the reconstructed 3D positions of the events. This FV was the result of an optimization study designed to maximize the signal-to-background ratio by minimizing contributions from surface backgrounds and external $\gamma$ rays. Regarding the data utilized, a portion was excluded to mitigate the impact of neutron-activated isotopes generated during the neutron calibration campaign preceding the radon calibrations, and to better constrain high-energy backgrounds. For the $^{212}$Pb analysis, only 1.1\,days of the full $^{220}$Rn calibration campaign and an energy region of interest (ROI) of [30, 620]\,keV were used. Similarly, for the $^{214}$Pb analysis, 8.5\,days of the full $^{222}$Rn calibration campaign and an ROI of [20, 1030]\,keV were selected.

\section{Analysis Framework}

The branching ratios of $^{212}\mathrm{Pb}$ and $^{214}\mathrm{Pb}$ were extracted via a binned maximum likelihood fit to the reconstructed energy spectra of the calibration. The data were binned with widths of 2\,keV and 5\,keV for the $^{212}\mathrm{Pb}$ and $^{214}\mathrm{Pb}$ spectra, respectively. The analysis was performed using a custom-built modeling and inference developed for high-energy nuclear physics studies in XENONnT, capable of incorporating detector response effects, signal efficiencies, and background components~\cite{Ferrari2025Radon}.

A Poisson likelihood fit was employed with Gaussian constraints~\cite{BAKER1984437} on nuisance parameters and background rates, and the resulting fit was evaluated using a Poisson Chi-Square goodness-of-fit test via \texttt{GOFEvaluation}~\cite{robert_hammann_2021_5626909} with a predefined 5\% rejection threshold. The constraints were derived from ancillary studies, as described below.

We modeled the $^{212}\mathrm{Pb}$ decay via three channels: the GS and two excited states EX1 and EX2 (238.6\,keV, 415.3\,keV). For $^{214}\mathrm{Pb}$, five channels were considered: the GS and EX1–EX4 (295.2\,keV, 351.9\,keV, 533.7\,keV, and 839.0\,keV). According to the LNHB, $^{214}\mathrm{Pb}$ possesses a fifth decay branch to the 797\,keV excited state of $^{214}\mathrm{Bi}$, with a branching fraction of 0.0196\%. This transition was omitted from the analysis due to low detection sensitivity. Theoretical shapes from~\cite{Mougeot} were used for the first-forbidden non-unique GS $\beta$ decays. Given the absence of publicly available theoretical calculations for the excited state transitions, we simulated their spectral shapes using the Geant4 toolkit, and applied LXe microphysics and detector effects. Geant4's Radioactive Decay Module models the decay using nuclear data from the Evaluated Nuclear Structure Data File~\cite{osti_1845010}. The simulated $\beta$ spectra are sampled from theoretical distributions based on Fermi’s three-body phase space, incorporating the Fermi function for Coulomb corrections. Ideally, these distributions should also include a shape factor $S$, to account for the degree of forbiddenness of the transition. However, this factor is not currently implemented for $^{212}\mathrm{Pb}$ and $^{214}\mathrm{Pb}$ decays within the Geant4 framework, which instead assumes an allowed transition shape ($S=1$)~\cite{Geant4PhysicsReferenceManual_BetaDecay}. For the ground state transitions, we gauged the impact of these inaccuracies by comparing theoretical shapes; however, a comprehensive systematic evaluation necessitates a more precise treatment involving input from the nuclear theory community. 

Backgrounds were modeled using a combination of data-driven and simulation-based templates. 
For $^{212}\mathrm{Pb}$, the SR0 science data were used as a background template, scaled to match the exposure of the calibration run. Additional components included $^{212}\mathrm{Bi}$ events, generated by high-energy $\gamma$ rays emitted in the decays of $^{212}\mathrm{Bi}$ nuclei positioned outside the instrumented LXe volume, $^{212}\mathrm{Pb}$ PTFE events, originating from $^{220}\mathrm{Rn}$ daughters plating out onto the TPC PTFE surfaces, and neutron-activated $^{129\text{m}}\mathrm{Xe}$ and $^{131\text{m}}\mathrm{Xe}$ monoenergetic signals.
For the latter, given the proximity of the $^{129\text{m}}\mathrm{Xe}$ and EX1 $\gamma$-ray energies, we performed an ancillary study to constrain their contribution, via Gaussian terms in the likelihood, by exploiting SR0 background data. This study revealed that the background contribution from neutron-activated $^{133}\mathrm{Xe}$ was consistent with zero; consequently, it was excluded from the fit model.

For the $^{214}\mathrm{Pb}$ study, instead, the SR1 background data were used, with additional contributions from $\gamma$ rays from $^{214}\mathrm{Bi}$ decaying outside LXe instrumented volume, $^{214}\mathrm{Pb}$ PTFE surfaces events, $^{129\text{m}}\mathrm{Xe}$, $^{131\text{m}}\mathrm{Xe}$, and $^{133}\mathrm{Xe}$. In contrast to the $^{212}\mathrm{Pb}$ analysis, the $^{214}\mathrm{Pb}$ signal does not shadow the neutron-activated backgrounds and thus no additional constraints were needed. 

Other $^{212}\mathrm{Pb}$ and $^{214}\mathrm{Pb}$ decay chain components were not accounted for as background contributions, as their reconstructed energies fall outside the ROI, or they have half-lives orders of magnitude longer than the measurement timescale, resulting in a subdominant rate.

The energy resolution and the energy reconstruction bias were modeled as described in~\cite{XENONnT_AP1}. The SR0 and SR1 analysis had separate detector response, extracted from calibration data and monoenergetic spectral lines. The parameters describing these models were included in the fit, with their uncertainties applied as constraints.

\section{Results}

The fitted $^{212}\mathrm{Pb}$ and $^{214}\mathrm{Pb}$ signal rates were used to derive the BRs of the relative $\beta$ decay processes, with their sum normalized to unity. 
In order to take into account the possible correlations between the signal rate parameters, their statistical uncertainties were evaluated from the full likelihood covariance matrix. 

Furthermore, in order to conservatively incorporate systematic uncertainty, the concurrent measurement of the branching ratios for both $^{212}\mathrm{Pb}$ and $^{214}\mathrm{Pb}$, via best fit parameter values, was repeated for different configurations of the analysis framework. Specifically, the binning and the FV were varied as well as the ROI and the data selection. The systematic uncertainty for each of these sources was then extracted as the maximum spread respect to the reference measurement. To obtain the total systematic uncertainty, all the uncertainties from the different sources were summed in quadrature. 

In order to probe the impact of a possible bias in the theoretical spectral-shape for the ground states, we reevaluated the fit results by employing their allowed shape versions~\cite{Mougeot}. The largest observed absolute difference between these results and the reference values corresponded to four times the statistical uncertainty. Given the qualitative nature of this check and the limited reliability of the allowed versions for the ground states, we decided not to include this estimate in the final systematic uncertainty.

\subsection{$^{212}\mathrm{Pb}$ Branching Ratios}

\begin{figure}
    \centering
    \includegraphics[width=0.9\linewidth]{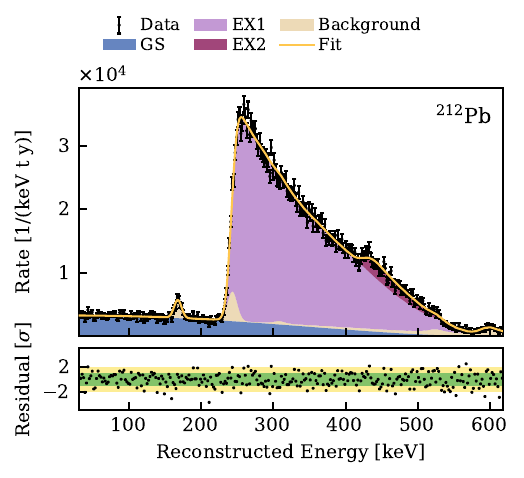}
    \caption{Results of the $^{212}\mathrm{Pb}$ BRs analysis. The best-fit model (yellow solid line) is plotted on top of the analyzed $^{220}\mathrm{Rn}$ calibration data. Signal fit model components (blue and purple shades) are stacked and non-signal contributions are grouped in the background template (sand). Normalized residuals are reported in the lower panel.}
    \label{fig:pb212model}
\end{figure}

The best-fit model is shown together with the $^{220}\mathrm{Rn}$ calibration data in Figure~\ref{fig:pb212model}. 
The goodness-of-fit test yields a p-value of 0.055. An inspection of the parameter pulls revealed no deviations exceeding $2\sigma$. 
Table~\ref{tab:pb212_branching} lists the resulting $^{212}\mathrm{Pb}$ BRs along with their associated statistical and systematic uncertainties.

\begin{table}[ht]
\centering
\setlength{\tabcolsep}{12pt} 
\renewcommand{\arraystretch}{1.5} 
\caption{Estimated $^{212}$Pb branching ratios.}
\label{tab:pb212_branching}
\begin{tabular}{l l}
\toprule
\textbf{State} & \textbf{Branching Ratio (\%)} \\
\midrule
GS  & $14.75 \pm 0.20 \text{\scriptsize (stat)}\,{}^{+0.14}_{-0.40} \text{\scriptsize (sys)}$ \\
EX1 & $80.4 \pm 0.3 \text{\scriptsize (stat)}\,{}^{+0.4}_{-0.4}  \text{\scriptsize (sys)}$ \\
EX2 & $4.9  \pm 0.3 \text{\scriptsize (stat)}\,{}^{+0.4}_{-0.5}  \text{\scriptsize (sys)}$ \\
\bottomrule
\end{tabular}
\end{table}

\begin{figure}
    \centering
    \includegraphics[width=1\linewidth]{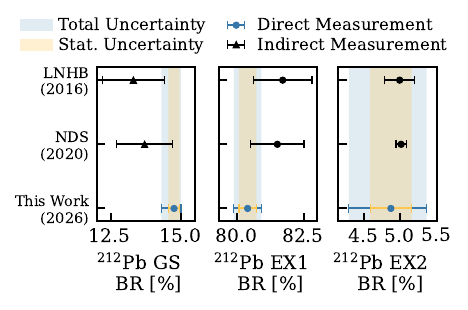}
    \caption{Branching ratios of $^{212}\mathrm{Pb}$ GS, EX1 and EX2 estimated by this work compared to the literature values. Measurements from this work are shown as light-blue points. Yellow (light-blue) error band indicate the statistical (total) uncertainties. LNHB~\cite{Be2016Table} and NDS~\cite{AURANEN2020117} values are reported in black.}
    \label{fig:pb212literature}
\end{figure}

Figure~\ref{fig:pb212literature} shows the comparison to literature values~\cite{Be2016Table,AURANEN2020117}. Our measurements are compatible within 2$\sigma$ with the two literature values, considering the systematic uncertainty.

\subsection{$^{214}\mathrm{Pb}$ Branching Ratios}

\begin{figure}
    \centering
    \includegraphics[width=1\linewidth]{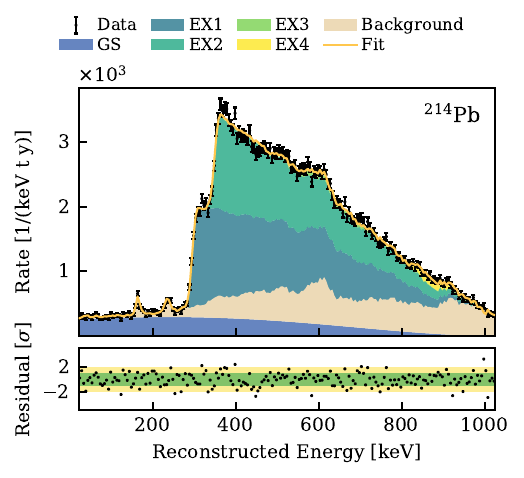}
    \caption{Results of the $^{214}\mathrm{Pb}$ BRs analysis. The best-fit model (yellow solid line) is plotted on top of the analyzed $^{222}\mathrm{Rn}$ calibration data. Signal fit model components (blue-green and yellow shades) are stacked and non-signal contributions are grouped in the background template (sand). Normalized residuals are reported in the lower panel.}
    \label{fig:pb214model}
\end{figure}

The result of the $^{214}\mathrm{Pb}$ analysis is reported in Figure~\ref{fig:pb214model}, where the best-fit model is shown on top of the analyzed $^{222}\mathrm{Rn}$ calibration data. 
The fit for $^{214}\mathrm{Pb}$ yielded a Poisson Chi-Square p-value of 0.074. No deviations on the parameter pulls above $2\sigma$ were observed. 
Table~\ref{tab:pb214_branching} summarizes the extracted $^{214}\mathrm{Pb}$ BR alongside their associated statistical and systematic uncertainties.

\begin{figure}
    \centering
    \includegraphics[width=1\linewidth]{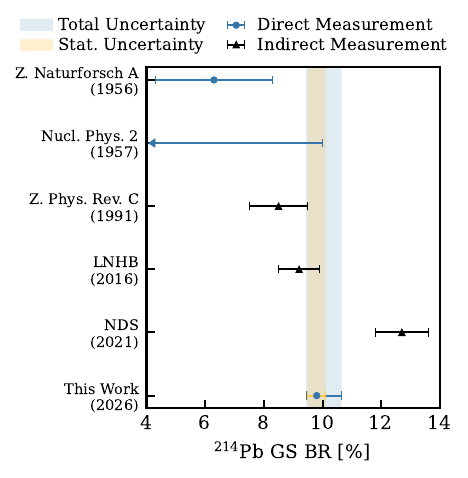}
    \caption{Comparison of this $^{214}\mathrm{Pb}$ GS BR measurement with literature reference values~\cite{Daniel_1956,NIELSEN1956476,ZHU20211,Be2016Table,PhysRevC.43.1639}. 
    Direct measurements, including this work, are reported with blue dots and blue total-uncertainty bars; upper limits are indicated by edge caps and arrows. Indirect measurements are reported with black triangles and black statistical uncertainty bars. The statistical and total uncertainties of this work are indicated by the yellow and blue vertical bands, respectively.}
    \label{fig:pb214literarure}
\end{figure}

\begin{table}[ht]
\centering
\setlength{\tabcolsep}{12pt} 
\renewcommand{\arraystretch}{1.5} 
\caption{Estimated $^{214}$Pb branching ratios.}
\label{tab:pb214_branching}
\begin{tabular}{l l}
\toprule
\textbf{State} & \textbf{Branching Ratio (\%)} \\
\midrule
GS  & $9.8 \pm 0.3 \text{\scriptsize (stat)}\,{}^{+0.8}_{-0.2} \text{\scriptsize (sys)}$ \\
EX1 & $42.6 \pm 0.8 \text{\scriptsize (stat)}\,{}^{+2.0}_{-1.9} \text{\scriptsize (sys)}$ \\
EX2 & $42.8 \pm 0.9 \text{\scriptsize (stat)}\,{}^{+0.8}_{-0.9} \text{\scriptsize (sys)}$ \\
EX3 & $1.9  \pm 0.7 \text{\scriptsize (stat)}\,{}^{+0.5}_{-1.7} \text{\scriptsize (sys)}$ \\
EX4 & $3.0  \pm 0.5 \text{\scriptsize (stat)}\,{}^{+0.9}_{-0.9} \text{\scriptsize (sys)}$ \\
\bottomrule
\end{tabular}
\end{table}
As shown in Figure~\ref{fig:pb214literarure}, the comparison between our $^{214}\mathrm{Pb}$ GS BR measurement and the literature reports a better agreement with LNHB~\cite{Be2016Table} and other direct measurements~\cite{Daniel_1956,NIELSEN1956476} than with the NDS value~\cite{ZHU20211}.
This result marks a step forward in solving the current LNHB-NDS literature $4.7\sigma$ tension for the $^{214}\mathrm{Pb}$ GS BR.

Concerning the excited states BRs resulting from this measurement, no strong preference among the two literature references is found. Moreover, given the small energy difference between EX1 and EX2 and the small predicted probabilities of accessing EX3 and EX4, the first two BRs are strongly correlated, while the second ones are affected by large uncertainties. Future studies, involving refined theoretical models and the integration of MS events in the analysis, will be able to reduce these correlations and uncertainties even further.

\section{Conclusions}

We reported high-precision direct measurements of the $\beta$ decay branching ratios of $^{212}\mathrm{Pb}$ and $^{214}\mathrm{Pb}$ using XENONnT radon calibration data. This study demonstrated that low-background LXe-TPCs, such as XENONnT, are highly suitable for precision nuclear physics measurements, reinforcing the results obtained in~\cite{b3r7-6ff4}. The experimental data were accurately described by our models, allowing us to verify the literature branching ratios for $^{212}\mathrm{Pb}$ decay and reduce the relative uncertainty of its ground-state branching ratio by a factor of three. Concerning $^{214}\mathrm{Pb}$, our measurement of the ground-state branching ratio agrees well with the LNHB evaluation~\cite{Be2016Table} and other direct measurements. Notably, we achieved a precision comparable to the LNHB evaluation, when accounting for systematic uncertainties. In a recent preprint currently under review, the PandaX collaboration~\cite{PandaX:2025azn} reported a GS transition value for $^{214}\mathrm{Pb}$ that is slightly higher than, but statistically consistent with, our measurement. Although that analysis utilized different theoretical spectral shapes for the excited states, the agreement remains within $2\sigma$, indicating that the approximate expressions for the excited-state spectral shapes in Geant4 do not have a dominant impact. These high-precision results are of significant importance to the broader nuclear physics community, providing crucial experimental benchmarks needed to improve and refine theoretical calculations of $\beta$ decay spectra.
Ultimately, these isotopes are the primary contributors to the low-energy ER background in LXe-TPCs. Because this background directly impacts the sensitivity of solar neutrino detection and searches for physics beyond the Standard Model, these results are critical for refining background constraints in both current and future analyses. Furthermore, they provide the precise inputs necessary to improve background simulations for next-generation observatories, such as XLZD~\cite{XLZD_design_book}, enabling a rigorous estimation of their ultimate science reach. Future improvements to this study may be achieved by adopting more sophisticated theoretical spectral shapes, incorporating increased statistics, and including multi-site data.

\begin{acknowledgments}

 We gratefully acknowledge support from the National Science Foundation, Swiss National Science Foundation, German Ministry for Education and Research, Max Planck Gesellschaft, Deutsche Forschungsgemeinschaft, Helmholtz Association, Dutch Research Council (NWO), Fundacao para a Ciencia e Tecnologia, Weizmann Institute of Science, Binational Science Foundation, Région des Pays de la Loire, Knut and Alice Wallenberg Foundation, Kavli Foundation, JSPS Kakenhi, JST FOREST Program, and ERAN in Japan, Tsinghua University Initiative Scientific Research Program, National Natural Science Foundation of China, Ministry of Education of China, DIM-ACAV+ Région Ile-de-France, and Istituto Nazionale di Fisica Nucleare. This project has received funding/support from the European Union’s Horizon 2020 and Horizon Europe research and innovation programs under the Marie Skłodowska-Curie grant agreements No 860881-HIDDeN and No 101081465-AUFRANDE.
\newline
\newline

We gratefully acknowledge support for providing computing and data-processing resources of the Open Science Pool and the European Grid Initiative, at the following computing centers: the CNRS/IN2P3 (Lyon - France), the Dutch national e-infrastructure with the support of SURF Cooperative, the Nikhef Data-Processing Facility (Amsterdam - Netherlands), the INFN-CNAF (Bologna - Italy), the San Diego Supercomputer Center (San Diego - USA) and the Enrico Fermi Institute (Chicago - USA). We acknowledge the support of the Research Computing Center (RCC) at The University of Chicago for providing computing resources for data analysis.

We thank the INFN Laboratori Nazionali del Gran Sasso for hosting and supporting the XENON project. 
\end{acknowledgments}

\bibliography{references}
\bibliographystyle{apsrev4-2}

\end{document}

%% file: affiliations.tex
\newcommand{\bologna}{\affiliation{Department of Physics and Astronomy, University of Bologna and INFN-Bologna, 40126 Bologna, Italy}}
\newcommand{\chicago}{\affiliation{Department of Physics, Enrico Fermi Institute \& Kavli Institute for Cosmological Physics, University of Chicago, Chicago, IL 60637, USA}}
\newcommand{\coimbra}{\affiliation{LIBPhys, Department of Physics, University of Coimbra, 3004-516 Coimbra, Portugal}}
\newcommand{\columbia}{\affiliation{Physics Department, Columbia University, New York, NY 10027, USA}}
\newcommand{\lngs}{\affiliation{INFN-Laboratori Nazionali del Gran Sasso and Gran Sasso Science Institute, 67100 L'Aquila, Italy}}
\newcommand{\mainz}{\affiliation{Institut f\"ur Physik \& Exzellenzcluster PRISMA$^{+}$, Johannes Gutenberg-Universit\"at Mainz, 55099 Mainz, Germany}}
\newcommand{\mpik}{\affiliation{Max-Planck-Institut f\"ur Kernphysik, 69117 Heidelberg, Germany}}
\newcommand{\munster}{\affiliation{Institut f\"ur Kernphysik, University of M\"unster, 48149 M\"unster, Germany}}
\newcommand{\nikhef}{\affiliation{Nikhef and the University of Amsterdam, Science Park, 1098XG Amsterdam, Netherlands}}
\newcommand{\nyuad}{\affiliation{New York University Abu Dhabi - Center for Astro, Particle and Planetary Physics, Abu Dhabi, United Arab Emirates}}
\newcommand{\purdue}{\affiliation{Department of Physics and Astronomy, Purdue University, West Lafayette, IN 47907, USA}}
\newcommand{\rice}{\affiliation{Department of Physics and Astronomy, Rice University, Houston, TX 77005, USA}}
\newcommand{\stockholm}{\affiliation{Oskar Klein Centre, Department of Physics, Stockholm University, AlbaNova, Stockholm SE-10691, Sweden}}
\newcommand{\subatech}{\affiliation{SUBATECH, IMT Atlantique, CNRS/IN2P3, Nantes Universit\'e, Nantes 44307, France}}
\newcommand{\torino}{\affiliation{INAF-Astrophysical Observatory of Torino, Department of Physics, University  of  Torino and  INFN-Torino,  10125  Torino,  Italy}}
\newcommand{\ucsd}{\affiliation{Department of Physics, University of California San Diego, La Jolla, CA 92093, USA}}
\newcommand{\wis}{\affiliation{Department of Particle Physics and Astrophysics, Weizmann Institute of Science, Rehovot 7610001, Israel}}
\newcommand{\zurich}{\affiliation{Physik-Institut, University of Z\"urich, 8057  Z\"urich, Switzerland}}
\newcommand{\paris}{\affiliation{LPNHE, Sorbonne Universit\'{e}, CNRS/IN2P3, 75005 Paris, France}}
\newcommand{\freiburg}{\affiliation{Physikalisches Institut, Universit\"at Freiburg, 79104 Freiburg, Germany}}
\newcommand{\napels}{\affiliation{Department of Physics ``Ettore Pancini'', University of Napoli and INFN-Napoli, 80126 Napoli, Italy}}
\newcommand{\nagoya}{\affiliation{Kobayashi-Maskawa Institute for the Origin of Particles and the Universe, and Institute for Space-Earth Environmental Research, Nagoya University, Furo-cho, Chikusa-ku, Nagoya, Aichi 464-8602, Japan}}
\newcommand{\laquila}{\affiliation{Department of Physics and Chemistry, University of L'Aquila, 67100 L'Aquila, Italy}}
\newcommand{\tokyo}{\affiliation{Kamioka Observatory, Institute for Cosmic Ray Research, and Kavli Institute for the Physics and Mathematics of the Universe (WPI), University of Tokyo, Higashi-Mozumi, Kamioka, Hida, Gifu 506-1205, Japan}}
\newcommand{\kobe}{\affiliation{Department of Physics, Kobe University, Kobe, Hyogo 657-8501, Japan}}
\newcommand{\kit}{\affiliation{Institute for Astroparticle Physics \& Institute of Experimental Particle Physics, Karlsruhe Institute of Technology, 76021 Karlsruhe, Germany}}
\newcommand{\tsinghua}{\affiliation{Department of Physics \& Center for High Energy Physics, Tsinghua University, Beijing 100084, P.R. China}}
\newcommand{\ferrara}{\affiliation{INFN-Ferrara and Dip. di Fisica e Scienze della Terra, Universit\`a di Ferrara, 44122 Ferrara, Italy}}
\newcommand{\groningen}{\affiliation{Nikhef and the University of Groningen, Van Swinderen Institute, 9747AG Groningen, Netherlands}}
\newcommand{\westlake}{\affiliation{Department of Physics, School of Science, Westlake University, Hangzhou 310030, P.R. China}}
\newcommand{\shenzhen}{\affiliation{School of Science and Engineering, The Chinese University of Hong Kong (Shenzhen), Shenzhen, Guangdong, 518172, P.R. China}}
\newcommand{\coimbrapoli}{\affiliation{Coimbra Polytechnic - ISEC, 3030-199 Coimbra, Portugal}}
\newcommand{\heidelberg}{\affiliation{Kirchhoff-Institute for Physics, Heidelberg University, 69120 Heidelberg, Germany}}
\newcommand{\roma}{\affiliation{INFN-Roma Tre, 00146 Roma, Italy}}
\newcommand{\bucknell}{\affiliation{Department of Physics \& Astronomy, Bucknell University, Lewisburg, PA, USA}}
\newcommand{\isct}{\affiliation{Department of Physics, School of Science, Institute of Science Tokyo, Meguro, Tokyo, 152-8551, Japan}}

%% file: authors.tex
\author{E.~Aprile\,\orcidlink{0000-0001-6595-7098}}\columbia
\author{J.~Aalbers\,\orcidlink{0000-0003-0030-0030}}\groningen
\author{K.~Abe\,\orcidlink{0009-0000-9620-788X}}\tokyo
\author{M.~Abu~Rmeileh\,\orcidlink{0009-0007-9750-6655}}\wis
\author{M.~Adrover\,\orcidlink{0123-4567-8901-2345}}\zurich
\author{S.~Ahmed~Maouloud\,\orcidlink{0000-0002-0844-4576}}\paris
\author{L.~Althueser\,\orcidlink{0000-0002-5468-4298}}\munster
\author{B.~Andrieu\,\orcidlink{0009-0002-6485-4163}}\paris
\author{E.~Angelino\,\orcidlink{0000-0002-6695-4355}}\lngs\chicago
\author{D.~Ant\'on~Martin\,\orcidlink{0000-0001-7725-5552}}\chicago
% ...
\collaboration{XENON Collaboration}\email[]{xenon@lngs.infn.it}\noaffiliation